\documentclass[copyright,a4paper,creativecommons]{eptcs}
\usepackage{amsmath}
\usepackage{amssymb}

\providecommand{\event}{FAVO 2009}
\providecommand{\volume}{??}
\providecommand{\anno}{20??}
\providecommand{\firstpage}{1}
\providecommand{\eid}{??}

\usepackage{breakurl}

\newcommand{\var}[1]{\ensuremath{\mathnormal{#1}}}
\newcommand{\term}[1]{\ensuremath{\langle\mathrm{#1}\rangle}}
\newcommand{\fn}[1]{\ensuremath{\mathrm{#1}}}
\newcommand{\definedby}{\ensuremath{\triangleq}}
\newcommand{\powerset}[1]{\ensuremath{\mathcal{P}(#1)}}

\newcommand{\secref}[1]{Section~\ref{sec:#1}}

\numberwithin{equation}{section}

\title{Common Representation of Information Flows\\for Dynamic Coalitions}

\author{
  I. Mozolevsky\quad and\quad J.S. Fitzgerald
  \institute{School of Computing Science\\Newcastle University\\UK}
  \email{\{igor.mozolevsky, john.fitzgerald\}@ncl.ac.uk}
}

\def\titlerunning{Common~Representation}
\def\authorrunning{I.~Mozolevsky~\&~J.~S.~Fitzgerald}

\begin{document}
\maketitle
\begin{abstract}
We propose a formal foundation for reasoning about access control
policies within a Dynamic Coalition, defining an abstraction over
existing access control models and providing mechanisms for
translation of those models into information-flow domain. The
abstracted information-flow domain model, called a Common
Representation, can then be used for defining a way to control the
evolution of Dynamic Coalitions with respect to information flow.
\end{abstract}

\section{Introduction}\label{sec:intro}\label{sec:dc}
A \textit{Dynamic Coalition} (DC) is a group of independent
organisations collaborating towards the achievement of a goal that the
member organisations could not achieve alone. The dynamic character
comes from the fact that membership, and the capabilities of members,
may change over time.  The heterogeneity of the DC membership,
bringing together organisations with very different policies governing
access to their services, gives rise to security concerns that would
normally not be associated with a well-defined static organisation. In
order to collaborate effectively, resources need to be shared between
the participants. Many DCs emerge through ad-hoc assembly but it is
nevertheless desirable to engineer infrastructure and policies to
enable aspects of the behaviour of coalitions to be verified in
advance of operation.

We treat a dynamic coalition as a group of autonomous communicating
agents~(the coalition members)~\cite{Bryans&06d,Bryans&07}. Each
member provides services that may be requested by other members and
which may entail the granting of access to resources.  Each
organisation that participates in a DC may be represented as a whole
as one member of the DC or as number of members. The level of
abstraction in this case depends on the desired level of atomicity of
an agent. For example, in the naval context an agent may represent the
whole of a vessel or merely the vessel's captain.

Much of the challenge in designing information flow policies and
infrastructure to support a DC lies in coping with the changing
membership with the need to potentially renegotiate access and
reallocate tasks from the workflow~\cite{Bryans&07}.

Ideally, the dynamic properties of membership should not disrupt the
operation of the coalition where both availability and confidentiality
are equally significant. These important requirements are often
overlooked and simplified~(at least for dynamically reconfigurable
systems) by, for example, assuming that the operations can be
suspended while the reconfiguration takes place
\cite{875265,10.1109/32.60317,271244}; in reality one cannot suspend
operations while adding new agents and assessing the impact of new
agents. Furthermore, as with most complex systems, security is often
added after the design stage as a bolt-on, which is unacceptable where
confidentiality and dependability of information is a
significant factor. The problem of maintaining security properties of
a dynamic coalition are further compounded by heterogeneity -- having
numerous agents that potentially have different and conflicting access
control policies. This gives rise to the need to be able to reason
about and assess the impact of dynamic changes on coalition
membership.

This paper focuses on security considerations in DC and proposes the
use of a \textit{Common Representation} (CR) that allows coalition
designers and architects to evaluate the effects of changes in
membership of a coalition and interaction between agents. The common
representation describes information flows between agents and
resources within a DC. Functions over the CR describe ways of
analysing the composition of numerous access control policies, the
impact of agents joining and departing a coalition, coalition merging,
and conflicts that may arise during a coalition's life.

By introducing the common representation, the paper addresses the
problem of managing access controls in dynamic coalitions where there
already exist access restrictions between subsets of agents without
having to design yet another access control model. Having a set of
algebraic expressions also allows coalition engineers to reason about
the liveliness (ability of the coalition to complete its goals) and
the confidentiality properties of different configurations of access
control restrictions.

The CR, in turn, is a foundation block of a Security Meta-Policy (SMP)
which is used to guide and control the evolution of a
DC by specifying security properties that must be preserved
every time there is a change in the coalition and by describing how
security-related changes must be implemented.

The paper consists of three parts. First, we first consider various
access control methods in use
today~(Section~\ref{sec:drawbacks}). Second, then we look at how the
common representation is expressed using directed
graphs~(Section~\ref{sec:cr}) and show how a selection of current
access control methods can be expressed in terms of this common
representation~(Section~\ref{sec:translation}). Finally we examine the
use of the common representation to define approaches to coalition
composition~(Sections~\ref{sec:comp} and~\ref{sec:props}). We end with
conclusions and a discussion of future work.

\section{Access Control in Coalitions}\label{sec:drawbacks}

As the information within a coalition is shared between various
agents, some of whom may be competitors, the ability to guard the
information against misuse and to prevent this information from
leaking uncontrollably from the coalition is of paramount importance.

For the purposes of the present study we treat each DC member as a
single entity. In practice, members may themselves be composed of
individuals that receive data and issue responses on behalf of the
member as a whole. We also assume that each individual tasks in the
workflow being executed by the DC may entail invocations a single
service by a single agent. We therefore assume that the workflow has
been broken down to a relatively low level of granularity. We revisit
these assumptions in Section~\ref{sec:concl}.

The most elementary form of access control is \textit{Discretionary
  Access Controls} (DACs). In a DAC model, either the owner or the
custodian, of information or resource has full discretion over who and
at what level (e.\ g.\ write, read) is permitted access to the
resource. While this form of access control may have its place in an
organisation, the risk of information misuse and leakage is too great
for DACs to be practically used in a DC.

There already exist numerous \textit{non-discretionary} access control
mechanisms that are used in organisations today. The non-discretionary
access controls are typically predefined and are managed
centrally. The two that are most prominently used are
\textit{Lattice-Based Access Controls} (LBACs) \cite{619980} and
\textit{Role-Based Access Controls} (RBACs) \cite{226710}.  LBACs are
very popular in military and other government organisations as this
model is compatible with the `need-to-know' security principle whereby
every participant and piece of information is classified based on set
criteria. While such an approach may be perfect for one organisation,
different organisations have different classification criteria; for
example, the US and the UK military have different clearance
levels. If LBACs were to be used in a DC, then all participants and
information would have to either be reclassified based on new common
criteria or each participating organisation would have to communicate
through a trusted agent that would be responsible for ensuring that
only permitted information can leave the organisation. Both of these
solutions are unacceptable, as the first one creates a huge overhead
for each new organisation wishing to join the coalition, and the
second one creates choke points which limits communication bandwidth
between the agents. Additionally, non-government organisations may be
unfamiliar with or not structured around LBACs, which introduces
another complication for interactions between government agencies and
private companies as the two are considered to have different
priorities with respect to the information security --- Clark and
Wilson argued \cite{ClarkWilson87} that military is concerned with
confidentiality, while for private companies integrity is more
important.

RBAC is a more flexible scheme than LBAC whereby access is granted
based on the role of the participant. When the agents were to interact
across organisation boundaries, new roles would have to be created in
each organisation based on the involvement of the interaction. RBACs
still need to be controlled through a centralised authorisation
system, which leads to unreasonably high administration overhead for
large systems, especially when the participants may not necessarily be
well-known to the administering authority. Furthermore, without
frequent reviews, in the real world, RBACs are known to have a
privilege accumulation problem, where participants retain privileges
that are no longer required and may lead to information leakage when
misused.

Freudenthal et~al.~\cite{Freudenthal02drbac} have attempted to address
some of the drawbacks of the RBACs in distributed environments through
\textit{Distributed Role-Based Access Controls} (dRBAC) as part of a
larger project on Distributed Coalition Infrastructure
\cite{disco}. While dRBAC is a significant step in terms of ensuring a
degree of trust and distributing the responsibility for creation of
roles, this approach relies on creating new roles for participants and
is still prone to privilege accumulation.

In the context of DCs, all of the above methods of access control
require conversion of agents' policies from one to another. Performing
such conversion will require a set amount of time during which the
coalition membership must remain constant; depending on the rate of
change of membership, such constraint may be unfeasible. On top of
that, none of the aforementioned access control methods address
security policy compatibility across the participants of dynamic
coalitions, and providing strong guarantees that information will not
traverse through organisational boundaries and end up in the wrong
hands is nearly impossible without such analysis.

There already exist security policy description languages like {\em
  Ponder} \cite{Lupu01theponder}, {\em EPAL}, {\em XACML}
\cite{1180378} and {\em Hyperproperties} \cite{1381245}. These
languages rely on the premise that the policies will be specified in a
particular description language and then translated to models or
implemented in a way that the agents understand. The approach taken in
the CR is the opposite --- the existing security policies are
abstracted and compared for compatibility.

While the above-mentioned access controls could work to a degree in
dynamic coalitions, there is still a great focus on who can access
what, instead of managing the information flow directly. The {\em
  Common Representation} that we propose in this paper is
intended to translate access controls from the domain of subjects and
objects to information-flow domain and analysing the impact of such
flows directly. The information-flow domain allows us to reason how
information is exchanged between agents in a coalition and what
information agents might be privy to.

\section{Common Representation}\label{sec:cr}
\textit{Common Representation} (CR) provides a fresh approach to a
policy-neutral method of defining access control constraints for a
dynamic system by focusing on the flow of information within that
system. In the context of dynamic coalitions, this represents the
information exchange between agents and resources. The CR also allows
coalition architects to reason about and analyse the effects of
permitting new interactions between agents on confidentiality and what
the impact departures of agents have on availability of information.

The robustness of CR stems from an observation that groups of agents
joining a coalition already have an access control policy guarding the
interaction between them. Thus, to provide secure communication
efficiently, merely providing a bridge between various groups of
agents would suffice, deferring all access control to run-time.
\subsection{Elements of a CR}
CR only concerns itself with \textit{read} and \textit{write}
privileges as these are the only ones that cause information to be
exchanged. If any other access control modes lead to data flows,
e.~g.~\textit{execute} privilege, those, too, can be refined down to
read and write operations.

A common representation is a directed graph \mbox{$\var{CR}\mathop=(I,
  F)$} where \textit{interfaces}, $I$, represent vertices and
\textit{flows}, \mbox{$F\subseteq I\times I$}, represent edges.

An interface is an abstract representation of a communication
port. There are two types of interfaces: \textit{explicit} and
\textit{implicit}. An explicit interface is an interface that
represents a tuple of a resource and an access mode,
e.~g.~$(r_1,\term{w})$; this type of interface an be either
\textit{active} (send/write) or \textit{passive} (receive/read). An
implicit interface can only represent an agent and can be both active
and passive. Implicit interfaces are an important utility in
modelling information flows where information is exchanged without a
dedicated (or perceived) resource, for example, in
face-to-face communication. Any agent can have numerous implicit
interfaces. The choice of what type of interface to use
largely depends on the desired level of abstraction required by the
security policy, although the decision has repercussions on the
operations as well as the ability to analyse resulting information
flows.

Flows in the $\var{CR}$ graph are used to represent permitted
information exchange paths between interfaces and are expressed as
pairs of interfaces. A read from $i_a$ to $i_b$ would be expressed as
$(i_a,i_b)$ while a write from $i_a$ to $i_b$ would be expressed as
$(i_b,i_a)$. Each flow is asymmetric and if a flow $(i_a,i_b)$ exists,
there is no implication of a complementary flow $(i_b,i_a)$. The flows
are always expressed left-to-right,
i.~e.\ $(i_{\var{from}},i_{\var{to}})$. Any flow expressed as
$(i_a,i_b)$, could mean either \mbox{$i_a\xleftarrow[from]{read}i_b$}
or \mbox{$i_b\xrightarrow[to]{write}i_a$}, which are equivalent. This
equivalence and asymmetry means that a CR will require two flows to
represent a bi-directional exchange of data:
$(i_a,i_b),(i_b,i_a)$. Bi-directional flows are called
\textit{complementary} --- two flows $f$ and $f^{-1}$ are
complementary if and only if $f\equiv(i_a,i_b)\wedge
f^{-1}\equiv(i_b,i_a)$.

Directed representation of flows has a notable side effect ---
``read'' and ``write'' lose their semantic meaning when converted into
information flows --- the only way that the original nature of a
particular flow would be preserved is if at least one of the
interfaces is explicit; this highlights the importance of not relying
on implicit flows.

Access, with respect to the access control policies, is an operation
which leads to information being exchanged between objects and
subjects. Thus by expressing the common representation as a
directional graph of information flows we can reason about information
access within the coalition.

The nature of any DC is that access can be granted or revoked on
several occasions. This level of abstraction allows explicit
interfaces to be destroyed and instantiated on-request (when an agent
leaves the coalition for example) thus fulfilling the demand for
dynamic changes in access controls to a resource. Additionally, each
agent can have its own dedicated interface to each of the resources
shared by other agents. Thus when an interface is destroyed because an
agent leaves, other agents can still retain access to the same
resources in the coalition.

Additionally, this representation allows any resource to have multiple
interfaces, a concept which could be used to model polyinstantiation
--- an explicit interface would be a third order object in the mapping
of $A\rightarrow R\rightarrow I$.
\subsection{Granting Access}
Securing the interaction between agents is based on the Clark-Wilson
model~\cite{ClarkWilson87} around a Trusted Computing Base (TCB)
described by Lampson et~al.~\cite{Lampson92authentication}. The
Clark-Wilson model refers to the notion of mediated access between an
object and a subject:
$\var{Subject}\longrightarrow\var{Mediator}\longrightarrow\var{Object}$
and TCB brings about various requirements, like hardware and software
support, for the mediator. The TCB refers to the mediator as a
\textit{reference monitor}, which in itself is an abstract
representation. The reference monitor consists of numerous
\textit{security kernels}. Each security kernel is made of trusted and
verified software and hardware. As these security kernels are verified
and are trusted to perform within specification, only the security
kernels are permitted to interact between agents. Therefore, removing
the need to rely on correct behaviour of any agent.

Access, by definition, is a process where information flow occurs from
$a$ to $b$. From this definition, granting access from one interface
to another within the CR is simply the matter of verifying whether a
particular flow is permitted.

In the absence of SMP and any other constraints, the default behaviour
for a security kernel for granting access between to interfaces can be
described by a simple function $\fn{grant}$. For brevity, in this
paper, we will assume that access will only be granted if the CR is
authoritative over both interfaces, otherwise the result of the
function is undefined. For a CR to be {\em authoritative} over an
interface, the interface must be defined in the vertices of the CR
graph.
\begin{equation*}
  \begin{split}
    {\mathrel{\fn{grant}}\colon}&\ I\times I\times \var{CR}\rightarrow\mathbb{B}\vert\mathbf{undef}\\
    \fn{grant}(&i_1,i_2,(i_s,f_s))
    \begin{cases}%
      (i_1,i_2)\in f_s & \text{if $\{i_1, i_2\}\subseteq i_s$}\\
      \mathbf{undef} & \text{otherwise}
    \end{cases}
  \end{split}
\end{equation*}

\section{Representing Basic Policies}\label{sec:translation}
The previous section gave an overview of the structure of common
representation. This section will discuss how common access control
policies like those described in Section~\ref{sec:drawbacks} are
mapped into common representation.
\subsection{Discretionary Access Controls}
Discretionary Access Controls are typically expressed as a
\textit{Capability Lists} or \textit{Access Control Lists} (ACL),
either of which is derived from an \textit{Access Control Matrix}
(ACM). An access control matrix is a table describing permitted
interactions between objects $(o_1,o_2,o_3)$ and subjects
$(s_1,s_2,s_3)$. An example ACM is shown in Table \ref{tab:acm}.

\begin{table}[!ht]
  \centering
  \begin{tabular}{r|c c c}
    &$o_1$&$o_2$&$o_3$\\
    \hline
    $s_1$&r&w&rw\\
    $s_2$&-&w&r\\
    $s_3$&rw&-&r
  \end{tabular}
  \caption{Simple Access Control Matrix}
  \label{tab:acm}
\end{table}

The access control matrix is a form of discretionary access control
where the permission mapping between subjects and objects remain
fairly static and well-defined at any one time. The next two access
controls fall into a non-discretionary access controls category where
instead of having a clear definition of interactions, access is
granted based on a set of predefined criteria.

The access control matrix shown in Table \ref{tab:acm} can be
described as an access control list:
\begin{equation*}
\begin{split}
\{&o_1\mapsto\left\{(s_1,\term{R}), (s_3,\term{R}), (s_3,
\term{W})\right\},\\
&o_2\mapsto\left\{(s_1,\term{W}), (s_2,\term{W})\right\},\\
&o_3\mapsto\left\{(s_1,\term{R}), (s_1,\term{W}), (s_3,\term{R})\right\}\}
\end{split}
\end{equation*}

For any ACM that describes access controls between a set of subjects,
$S$, and a set of objects, $O$, the DAC can be translated to a CR
containing $O\cup S$ as the set of interfaces. The flows of the CR are
based on the {\em read}- and {\em write}- operations permitted by the
access controls. ACLs are expressed as mappings from
objects to a set of subjects and access modes those subjects have to a
particular object: $\mathrel{\var{ACL}}\colon\
O\rightarrow\powerset{S\times M}$. Translating an ACL to CR once
objects and subjects have been mapped to interfaces is only a matter
of determining resulting 
\begin{equation*}
  \begin{split}
    &\left\{((s,\term{R}), (o,\term{W}))\vert o\in O, (s,\term{W}) \in
    \var{ACL}(o)\right\}\\
    &\qquad\cup\\
    &\left\{((o,\term{R}), (s, \term{W}))\vert o\in O, (s,\term{R})\in \var{ACL}(o)\right\}
  \end{split}
\end{equation*}
\noindent
Capability Lists can be translated into common representation in a
similar manner.
\subsection{Lattice-Based Access Control}
A Lattice-Based Access Control policy typically consists of three
components: subjects, objects and labels. Each subject and object is
marked with a label, then the labels are organised into a
partially ordered dominance hierarchy $(L, \leq_L)$, a special
abstract function $\mathrel{\lambda}\colon S\cup O\rightarrow L$ is then
used to determine the label associated with each object or subject.

The most commonly used LBAC policy is the Bell-LaPadula security model
\cite{blp}. The underlying principle of the model is that access is
determined by the dominance relationship between the object and the
subject and that the subjects and objects are interchangeable for the
purpose of modelling. The model states that $A$ can write to $B$ and
$B$ can read from $A$ if $\lambda(A)\leq_L\lambda(B)$. This model
reflects the equivalence of flows in the common representation that
was discussed earlier.

The derivation of the common representation for LBACs is similar that
for DACs except that read- and write- flows are treated
differently. Let us assume that set $I$ is a union of $S$ and
$O$. Total flows are derived as
\begin{equation*}
    \left\{(i_1,i_2)\vert i_1,i_2\in I\cdot i_1\neq
    i_2\wedge\lambda(i_1)\leq_L\lambda(i_2)\right\}
\end{equation*}
\subsection{Role-Based Access Controls}
Role-Based Access Controls differ from the access controls described
above in that there the relationship between objects and subjects is
not defined within the access control, instead, the subject acquires
privileges based on the role assignment. The implication of this
phenomenon is that while it remains possible to derive the flows of
information within each role across different roles, the
representation of information exchange between subjects is deferred
until the subject is assigned a particular role. Additionally, there
is a notion of role seniority, where a senior role absorbs permissions
of junior roles associated with it.

There are two components that are involved in expressing an RBAC: role
assignments and role hierarchies. The role assignments consist of a
mapping from a role to the privileges that role grants:
\mbox{$\mathrel{\var{RA}}:\var{R}\rightarrow\powerset{O\times
    M}$}. The role hierarchy is represented with the role seniority
relationship:
\mbox{$\mathrel{\var{RH}}\subseteq\var{R}\times\var{R}$}. Due to this
seniority relationship, it is necessary to determine transitive closure
of each role described by the hierarchy. This can be done using
Warshall's algorithm \cite{321107} or a similar algorithm described by
Osborn \cite{Osborn02informationflow}, which is more closely related
to role hierarchies; let the transitive closure over $\var{RH}$ be
$\var{RH}^{+}$. The mapping of senior roles,
\mbox{$\mathrel{\var{seniority}}:\var{R}\rightarrow\powerset{R}$}, to
junior roles can then be obtained through:
\begin{equation*}
        \mathrel{\var{seniority}}=\{r_j\vert(r,r_j)\in\var{RH}^+\}
\end{equation*}
we can then go on to determine what privileges a specific role gives
using \mbox{$\mathrel{\var{privileges}}\colon\var{R}\rightarrow\powerset{O\times\var{Mode}}$}:
\begin{equation*}
  \mathrel{\var{privileges}}=\bigcup\{\var{RA}(r_j)\vert r_j\in\var{seniority}(r)\}\cup\var{RA}(r)
\end{equation*}
and finally, we can derive the flows that arise in the RBAC based
on roles and the role hierarchy:
\begin{equation*}
 \mathrel{\var{flows}}=\bigcup\left\{\left\{((o,\term{R}),(o,\term{W}))\vert(o,\term{R}),(o,\term{W})\in\var{privileges}(r)\right\}\vert
 r\in R\right\}
\end{equation*}

\section{Compositionality of CRs}\label{sec:comp}
Having looked at how individual access control policies can be
translated into the common representation, we can now look at how
these common representations can be combined together to form an
overall common representation of information flow for the whole of a
dynamic coalition.

All operations on CRs assume that the names of the interfaces in all
CRs, which are involved in a particular operation, are named
consistently. That is, if $\var{CR_1}$ contained interface named
$\var{i_a}$ and $\var{CR_2}$ contained interface named $\var{i_a}$,
then both CRs refer to the same, identical interface. 

There are two composites that can be formed from a collection of
common representations - the {\em Simple Composite} and the {\em
  Priority Composite}. Both operations have equal precedence. Thus it
is important to explicitly group expressions containing both of these
composites.
\subsection{Simple Composite \texorpdfstring{---}{-} Merge}
When two groups of agents join to form a coalition, their respective
common representations need to be merged together to form a
coalition-wide representation of information flows. The most likely
scenario for a coalition formation is when the agents in separate
groups are isolated from other groups in terms of access
privileges. This merge is achieved by performing a simple composite.

The simple composite is the most straightforward way of combining
numerous common representations. This composite favours {\em
  permit}-type flows and will effectively result in a less strict
policy overall compared to any individual common representation.
\begin{equation*}
  \begin{split}
    \mathrel{\fn{merge}}\colon&\ \var{CR}\times\var{CR}\rightarrow\var{CR}\\
    \mathrel{\fn{merge}}(&(i_a,f_a),(i_b,f_b))\definedby(i_a\cup i_b, f_a\cup f_b)
  \end{split}
\end{equation*}
This {\em merge} operation will result in a graph of order two or
higher, if the two merging CRs do not have any interfaces in common,
as a case with independent 
All simple composites have commutative properties when applied as a
distributed operation over a set of common representations.
\subsection{Priority Composite \texorpdfstring{---}{-} Append}
Numerous times during the evolution of a dynamic coalition, a group of
unassociated agents will inevitably join an existing coalition. When a
group of agents joins in, the flow restrictions defined by the
established coalition must be respected, thus any group that is
merging-in will only create flows to the additional agents. This
type of merging is called a priority composite, with priority given to
one common representation.  The priority composite is a way of
combining common representations which favours {\em deny} over
permit. The priority composite will not allow new flows to be created
for {\em existing} interfaces where there does not exist a flow
already.
\begin{equation*}
  \begin{split}
    \mathrel{\fn{append}}:\ &\var{CR}\times\var{CR}\rightarrow\var{CR}\\
    \fn{append}(&(i_a,f_a),(i_b,f_b))\definedby(i_a\cup i_b, f_a\cup\{f\vert f\in f_b\cdot\{f,f^{-1}\}\cap f_a=\emptyset\})
  \end{split}
\end{equation*}
The priority composite is a non-commutative operation and operates
over an ordered set of common representations with a left-to-right
order.

\section{Analysing CRs and Composite Properties}\label{sec:props}
In this section we will look at how CRs can be analysed and look at
how we might determine conflicts.

We can use a CR to determine the degree of availability of information
flow paths within a DC using simple graph theory. For example, in
determining whether there is a potential for information to be
exchanged between two agents, we can check whether there exists a walk
between the relevant interface of said agents. This is an
important aspect when determining whether a particular interface is
critical to the operation of the DC. We can derive the availability
graph $\var{CR^A}$ of $\var{CR}$ by replacing all complementary flows
in $\var{CR}$ with an edge in $\var{CR^A}$. The availability graph is
an undirected graph and serves the purpose of determining whether
information can flow between the interfaces of the DC. If the number
of components in $\var{CR^A}$ is equal to one then we say that the
{\em liveliness property} of a DC holds.

We will now proceed to look at conflicts in CRs. A {\em CR conflict}
is said to occur when one common representation contains a flow
between any two interfaces, and the second one does not contain a flow
between the {\em corresponding interfaces}.
\subsection{Identifying Conflicts}
The first step to analysing any composite is to be able to identify
conflicts between different common representations.
\begin{equation}
  \begin{split}
    \mathrel{\fn{conflicting}}:\ &\var{CR}\times\var{CR}\rightarrow\mathbb{B}\\
    \fn{conflicting}(&(i_a,f_a),(i_b,f_b))\definedby \exists i_1,i_2 \in
    (i_a\cap i_b) \cdot i_1 \neq i_2 \wedge (i_1,i_2)\not\in
    (f_a\cap f_b)
  \end{split}
\end{equation}
\subsection{Analysing Conflicts}
We can further expand on conflict resolution by zooming in on specific
properties of CRs. For example, we can identify all flows that are
permitted in interfaces $a$ but denied in $b$ are determined by
expression in \eqref{eq:cflt}.  Note that, in this case, a conflict
may only arise if both CRs include identical interfaces. For example,
given two CRs
\begin{equation*}
  \var{CR_1}=(\{i_a,i_b, i_c\}, \{(i_a,i_c),
  (i_b,i_c)\})
\end{equation*}
and
\begin{equation*}
\var{CR_2}=(\{i_a,
  i_c,i_d\},\{(i_a,i_d),(i_d,i_c)\})
\end{equation*}
the only conflict that arises is \mbox{$(i_a,i_c)$}, as $i_b$ is not a
vertex in $\var{CR_2}$ and $i_d$ is not a vertex in $\var{CR_1}$.
\begin{equation}
  \begin{split}
    \mathrel{\fn{conflicts}}\colon&\var{CR}\times\var{CR}\rightarrow
    F\\ \fn{conflicts}(&(i_a,f_a),(i_b,f_b))\definedby\\&\{(i_1,
    i_2)\vert\forall i_1, i_2 \in (i_a \cap i_2) \cdot i_1\neq i_2
    \wedge (i_1,i_2)\not\in (f_a\cap f_b)\}
  \end{split}
  \label{eq:cflt}
\end{equation}
Similarly, flows common to both $a$ and $b$ can be identified from the
intersection of edges of CR graphs $a$ and $b$. Finally, the flows
that are different to both $a$ and $b$:
\begin{equation}
  \begin{split}
    \mathrel{\fn{diffs}}:\ &\var{CR}\times\var{CR}\rightarrow F\\
    \fn{diffs}(&(\_,f_a),(\_,f_b))\definedby(f_a\cup f_b)-(f_a\cap f_b)
  \end{split}
\end{equation}
\subsection{Compositionality Policies}
With the aid of algebraic operations described above, coalition
architects can define compositionality policies which can then be used
during the lifetime of the coalition to automate the process of
assimilating groups of agents into the coalition without having to
suspend operations while policies are integrated. Such policy may
state, for example, ``\textit{Conflicting policies \textbf{A} and
  \textbf{B} may be combined if the conflicts are limited to
  complementary flows of \textbf{A}, otherwise, the composite policy
  must not create extra permissions than those already defined by
  \textbf{A}}''. This can be expressed as:
\begin{equation*}
  \begin{split}
    \mathbf{if}\ &\forall f\in \fn{conflicts}(A,B)\cdot\exists
    f'\in\mathrm{flows}(A)\cdot
    f'=f^{-1}\\&\mathbf{then}\ \fn{merge}(A,B)\\&\mathbf{else}\ \fn{append}(A,B)
  \end{split}
\end{equation*}
\noindent 
where $\mathrm{flows}$ is an abstract function determining information
flows within a graph.

Defining policies like these is one of the purposes for the Security
Meta-Policy framework which takes the primitives of access controls
and security policies and describes ways to constrain them and perform
operations on policy-neutral representations in order to expedite the
integration of new agents, impose coalition-wide constraints on
policies and provide a mechanism for evaluating the effects of policy
manipulation.

\section{Conclusion and Future Work}\label{sec:concl}
In this paper we looked at a novel way of modelling security using a
CR, how a CR is structured and discussed ways of representing various
access control policies in the information-flow domain. We also have
demonstrated how a CR can be used to address {\em availability} and
{\em confidentiality} of the CIA security triad. Once the policies are
expressed in CR, we looked at various operations for composition and
analysis of policies.

The CR preserves the access constraints expressed in access control
and security policies by design and compositional algebra, provides a
way of manipulating not only different, but also conflicting
policies. While this approach is sufficient for simple problems,
dealing with more advanced concepts, like Separation of Duty or
Brewer-Nash Models, requires the use of additional layer of
parametrisation over the CR as well as the ability to define conditional
evaluation operations. This layer is provided by the Security
Meta-Policy Framework which is an ongoing work. The CR is
put to use in the Security Meta Policy where the CR expression algebra
is used for determining effects of changes in dynamic coalitions.

When the CR combined with a Security Meta-Policy, there exists a
possibility to group interfaces by semantic meaning together to form
blocks, modifying the definition of $\var{CR}$ to
\mbox{$\mathrel\var{CR}:(B,F)$} where $B\subseteq\powerset{I}$ and
$F\subseteq\powerset{B\times B}$. Each block can represent, for
example, a set of interfaces that have unrestricted flow of
information between each other, or a set of interfaces where flows are
forbidden, or where the number of flows is limited between the
interfaces from a block to any other block. This also allows us to
model interactions that were previously not possible in other access
control methods, like groups of people.

Additionally, given that CR is essentially a graph, it should be
possible to define logic for optimisation of flows, thus reducing the
number of unnecessary edges which in turn enforces the principle of
least privileges. Another way of optimising the CR would be to group
various complimentary interfaces together into individual blocks
(enabled by the use of SMP). This approach would reduce the complexity
of operations but would have a side effect of limiting the ability to
precisely express details of interaction.

Coming back to the assumptions made in Section~\ref{sec:drawbacks}, we
can now see how the atomicity of agents and granularity of tasks would
affect the structure of the CR. However, this should not be a problem
when translating existing security policies into CR, as the desired
levels of atomicity and task granularity should already be pre-defined
in the source policies. When new tasks are created, or agents added,
and a new CR is generated to include the desired changes, the new CR
can be checked for conflicts against the existing CR using techniques
described in \secref{props}. Results of such checks can be used to
gauge whether there needs to be a further division of tasks or whether
a particular flow is too permissive. This logic can also be included
into the SMP to automate the decision making process and change
configuration of the DC on-the-fly.

Finally, work on CRs could be linked with work on insider threat
modelling described by Chinchani et al
\cite{Chinchani05,Chinchani07}. These models, at the most fundamental
level, rely on interaction graphs between different nodes of network
to determine what node is exposed to what information. Such approach
is similar to what common representation provides, thus by moving
elements of insider threat modelling into the common representation,
the risk of insider abuse could be analysed and mitigated.

\paragraph{Acknowledgements:}
We are grateful to Anirban Bhattacharyya, Jeremy Bryans, Peter Ryan
and Alexander Romanovsky for many helpful discussions and valuable
insights.

\bibliographystyle{eptcs}
\bibliography{favo}

\begin{thebibliography}{10}
\providecommand{\bibitemstart}[1]{\bibitem{#1}}
\providecommand{\bibitemend}{}
\providecommand{\bibliographystart}{}
\providecommand{\bibliographyend}{}
\providecommand{\url}[1]{\texttt{#1}}
\providecommand{\urlprefix}{Available at }
\providecommand{\bibinfo}[2]{#2}
\bibliographystart

\bibitemstart{875265}
\bibinfo{author}{J.~P.~A. Almeida}, \bibinfo{author}{M.~van Sinderen} \&
  \bibinfo{author}{L.~Nieuwenhuis} (\bibinfo{year}{2001}):
  \emph{\bibinfo{title}{Transparent Dynamic Reconfiguration for CORBA}}.
\newblock In: {\sl \bibinfo{booktitle}{DOA '01: Proceedings of the Third
  International Symposium on Distributed Objects and Applications}},
  \bibinfo{publisher}{IEEE Computer Society}, p. \bibinfo{pages}{197}.
\bibitemend

\bibitemstart{1180378}
\bibinfo{author}{A.~H. Anderson} (\bibinfo{year}{2006}):
  \emph{\bibinfo{title}{A comparison of two privacy policy languages: EPAL and
  XACML}}.
\newblock In: {\sl \bibinfo{booktitle}{SWS '06: Proceedings of the 3rd ACM
  workshop on Secure web services}}, \bibinfo{publisher}{ACM}, pp.
  \bibinfo{pages}{53--60}.
\bibitemend

\bibitemstart{blp}
\bibinfo{author}{D.~Elliot Bell} \& \bibinfo{author}{Leopard~J. LaPadula}
  (\bibinfo{year}{1973}): \emph{\bibinfo{title}{Secure Computer Systems:
  Mathematical Foundations}}.
\newblock \bibinfo{type}{Technical Report} \bibinfo{number}{TR-2547},
  \bibinfo{institution}{MITRE Corporation}.
\bibitemend

\bibitemstart{Bryans&06d}
\bibinfo{author}{J.~W. Bryans}, \bibinfo{author}{J.~S. Fitzgerald},
  \bibinfo{author}{C.~B. Jones} \& \bibinfo{author}{I.~Mozolevsky}
  (\bibinfo{year}{2006}): \emph{\bibinfo{title}{{Formal Modelling of Dynamic
  Coalitions, with an Application in Chemical Engineering}}}.
\newblock In: \bibinfo{editor}{T.~Margaria}, \bibinfo{editor}{A.~Philippou} \&
  \bibinfo{editor}{B.~Steffen}, editors: {\sl \bibinfo{booktitle}{{IEEE-ISoLA
  2006: Second International Symposium on Leveraging Applications of Formal
  Methods, Verification and Validation}}}, pp. \bibinfo{pages}{90--97}.
\bibitemend

\bibitemstart{Bryans&07}
\bibinfo{author}{J.~W. Bryans}, \bibinfo{author}{J.~S. Fitzgerald} \&
  \bibinfo{author}{P.~Periorellis} (\bibinfo{year}{2007}):
  \emph{\bibinfo{title}{{A Formal Approach to Dependable Evolution of Access
  Control Policies in Dynamic Collaborations}}}.
\newblock In: {\sl \bibinfo{booktitle}{{Proc. 37th Annual IEEE/IFIP Intl. Conf.
  on Dependable Systems and Networks}}}, pp. \bibinfo{pages}{352--353}.
\bibitemend

\bibitemstart{Chinchani05}
\bibinfo{author}{R.~Chinchani}, \bibinfo{author}{A.~Iyer},
  \bibinfo{author}{H.~Ngo} \& \bibinfo{author}{S.~Upadhyaya}
  (\bibinfo{year}{2005}): \emph{\bibinfo{title}{Towards a Theory of Insider
  Threat Assessment}}.
\newblock In: {\sl \bibinfo{booktitle}{Proceedings of the International
  Conference on Dependable Systems and Networks}}, pp.
  \bibinfo{pages}{108--117}.
\bibitemend

\bibitemstart{ClarkWilson87}
\bibinfo{author}{D.~D. Clark} \& \bibinfo{author}{D.~R. Wilson}
  (\bibinfo{year}{1987}): \emph{\bibinfo{title}{A Comparison of Commercial and
  Military Computer Security Policies}}.
\newblock In: {\sl \bibinfo{booktitle}{IEEE Symposium on Security and
  Privacy}}, \bibinfo{publisher}{IEEE Computer Society}, p.
  \bibinfo{pages}{184}.
\bibitemend

\bibitemstart{1381245}
\bibinfo{author}{Michael~R. Clarkson} \& \bibinfo{author}{Fred~B. Schneider}
  (\bibinfo{year}{2008}): \emph{\bibinfo{title}{Hyperproperties}}.
\newblock In: {\sl \bibinfo{booktitle}{CSF '08: Proceedings of the 2008 21st
  IEEE Computer Security Foundations Symposium}}, \bibinfo{publisher}{IEEE
  Computer Society}, \bibinfo{address}{Washington, DC, USA}, pp.
  \bibinfo{pages}{51--65}.
\bibitemend

\bibitemstart{Lupu01theponder}
\bibinfo{author}{N.~Damianou}, \bibinfo{author}{N.~Dulay},
  \bibinfo{author}{E.~Lupu} \& \bibinfo{author}{M.~Sloman}
  (\bibinfo{year}{2001}): \emph{\bibinfo{title}{The Ponder Policy Specification
  Language}}.
\newblock In: {\sl \bibinfo{booktitle}{Lecture Notes in Computer Science}},
  \bibinfo{publisher}{Springer-Verlag}, pp. \bibinfo{pages}{18--38}.
\bibitemend

\bibitemstart{Freudenthal02drbac}
\bibinfo{author}{Eric Freudenthal}, \bibinfo{author}{Edward Keenan} \&
  \bibinfo{author}{Vijay Karamcheti} (\bibinfo{year}{2002}):
  \emph{\bibinfo{title}{dRBAC: Distributed role-based access control for
  dynamic coalition environments}}.
\newblock In: {\sl \bibinfo{booktitle}{In Proceedings of the Twenty-second IEEE
  International Conference on Distributed Computing Systems (ICDCS)}}, pp.
  \bibinfo{pages}{411--420}.
\bibitemend

\bibitemstart{disco}
\bibinfo{author}{F.~Freudenthal}, \bibinfo{author}{E.~Keenan},
  \bibinfo{author}{T.~Pesin}, \bibinfo{author}{L.~Port} \&
  \bibinfo{author}{V.~Karamcheti} (\bibinfo{year}{2001}):
  \emph{\bibinfo{title}{Dis{C}o: {A} {D}istributed {I}nfrastructure for
  {S}ecurely {D}eploying {D}ecomposable {S}ervices in {P}artially {T}rusted
  {E}nvironments.}}
\newblock \bibinfo{type}{Technical Report} \bibinfo{number}{2001-820},
  \bibinfo{institution}{Computer Science, New York University}.
\bibitemend

\bibitemstart{Chinchani07}
\bibinfo{author}{D.~Ha}, \bibinfo{author}{S.~Upadhyaya},
  \bibinfo{author}{H.~Ngo}, \bibinfo{author}{S.~Pramanik},
  \bibinfo{author}{R.~Chinchani} \& \bibinfo{author}{S.~Mathew}
  (\bibinfo{year}{2007}): \emph{\bibinfo{title}{Insider Threat Analysis using
  Information-Centric Modelling}}.
\newblock In: {\sl \bibinfo{booktitle}{Advances in Digital Forensics III}},
  \bibinfo{publisher}{Springer}, pp. \bibinfo{pages}{55--73}.
\bibitemend

\bibitemstart{10.1109/32.60317}
\bibinfo{author}{J.~Kramer} \& \bibinfo{author}{J.~Magee}
  (\bibinfo{year}{1990}): \emph{\bibinfo{title}{The Evolving Philosophers
  Problem: Dynamic Change Management}}.
\newblock {\sl \bibinfo{journal}{IEEE Transactions on Software Engineering}}
  \bibinfo{volume}{16}(\bibinfo{number}{11}), pp. \bibinfo{pages}{1293--1306}.
\bibitemend

\bibitemstart{Lampson92authentication}
\bibinfo{author}{B.~Lampson}, \bibinfo{author}{M.~Abadi},
  \bibinfo{author}{M.~Burrows} \& \bibinfo{author}{E.~Wobber}
  (\bibinfo{year}{1992}): \emph{\bibinfo{title}{Authentication in distributed
  systems: Theory and practice}}.
\newblock {\sl \bibinfo{journal}{ACM Transactions on Computer Systems}}
  \bibinfo{volume}{10}, pp. \bibinfo{pages}{265--310}.
\bibitemend

\bibitemstart{Osborn02informationflow}
\bibinfo{author}{S.~L. Osborn} (\bibinfo{year}{2002}):
  \emph{\bibinfo{title}{Information flow analysis of an {RBAC} system}}.
\newblock In: {\sl \bibinfo{booktitle}{7th ACM Symposium on Access Control
  Models and Technologies}}, \bibinfo{publisher}{ACM Press}, pp.
  \bibinfo{pages}{163--168}.
\bibitemend

\bibitemstart{226710}
\bibinfo{author}{R.~S. Sandhu}, \bibinfo{author}{E.~J. Coyne},
  \bibinfo{author}{H.~L. Feinstein} \& \bibinfo{author}{C.~E. Youman}
  (\bibinfo{year}{1996}): \emph{\bibinfo{title}{Role-Based Access Control
  Models}}.
\newblock {\sl \bibinfo{journal}{Computer}}
  \bibinfo{volume}{29}(\bibinfo{number}{2}), pp. \bibinfo{pages}{38--47}.
\bibitemend

\bibitemstart{619980}
\bibinfo{author}{Ravi~S. Sandhu} (\bibinfo{year}{1993}):
  \emph{\bibinfo{title}{Lattice-Based Access Control Models}}.
\newblock {\sl \bibinfo{journal}{Computer}}
  \bibinfo{volume}{26}(\bibinfo{number}{11}), pp. \bibinfo{pages}{9--19}.
\bibitemend

\bibitemstart{271244}
\bibinfo{author}{D.~B. Stewart}, \bibinfo{author}{R.~A. Volpe} \&
  \bibinfo{author}{P.~K. Khosla} (\bibinfo{year}{1997}):
  \emph{\bibinfo{title}{Design of Dynamically Reconfigurable Real-Time Software
  Using Port-Based Objects}}.
\newblock {\sl \bibinfo{journal}{IEEE Trans. Softw. Eng.}}
  \bibinfo{volume}{23}(\bibinfo{number}{12}), pp. \bibinfo{pages}{759--776}.
\bibitemend

\bibitemstart{321107}
\bibinfo{author}{S.~Warshall} (\bibinfo{year}{1962}): \emph{\bibinfo{title}{{A
  Theorem on Boolean Matrices}}}.
\newblock {\sl \bibinfo{journal}{Journal of the ACM}}
  \bibinfo{volume}{9}(\bibinfo{number}{1}), pp. \bibinfo{pages}{11--12}.
\bibitemend

\bibliographyend
\end{thebibliography}
\end{document}